\documentclass[pra,twocolumn,superscriptaddress,10pt]{revtex4-1}
\usepackage{graphicx}
\usepackage{dcolumn}
\usepackage{color}
\usepackage{times}
\usepackage{bm}
\usepackage{amssymb}
\usepackage{amsmath}
\usepackage{epsfig}
\usepackage{epstopdf}
\usepackage{dsfont}
\usepackage{subfigure}
\usepackage[colorlinks=true,citecolor=blue, linkcolor=blue]{hyperref}
\begin{document}
\renewcommand{\thefootnote}{\fnsymbol {footnote}}

\title{Improved tripartite  uncertainty relation with quantum memory}

\author{Fei Ming}
\affiliation{School of Physics \& Material Science, Anhui University, Hefei
230601,  People's Republic of China}

\author{Dong Wang} \email{dwang@ahu.edu.cn}
\affiliation{School of Physics \& Material Science, Anhui University, Hefei
230601,  People's Republic of China}
\affiliation{CAS  Key  Laboratory  of
Quantum  Information,  University  of  Science  and
Technology of China, Hefei 230026,  People's Republic of China}

\author{Xiao-Gang Fan}
\affiliation{School of Physics \& Material Science, Anhui University, Hefei
230601,  People's Republic of China}

\author{Wei-Nan Shi}
\affiliation{School of Physics \& Material Science, Anhui University, Hefei
230601,  People's Republic of China}


\author{Liu Ye} 
\affiliation{School of Physics \& Material Science, Anhui University, Hefei
230601,  People's Republic of China}

\author{Jing-Ling Chen} \email{chenjl@nankai.edu.cn}
\affiliation{Theoretical Physics Division, Chern Institute of Mathematics, Nankai University, Tianjin 300071, People's Republic of China}

\date{\today}

\begin{abstract}{
Uncertainty principle is a striking and fundamental feature in quantum mechanics distinguishing from classical mechanics.
It offers an important lower bound to predict outcomes of two arbitrary incompatible observables measured on a  particle.  In quantum information theory, this uncertainty principle is popularly formulized in
terms of entropy.
Here, we present an improvement of tripartite quantum-memory-assisted entropic uncertainty relation.
The uncertainty's lower bound is derived by considering mutual information and Holevo quantity.
It shows that the bound derived by this method will be tighter than
the lower bound in [Phys. Rev. Lett. 103, 020402 (2009)].
Furthermore, regarding a pair of mutual unbiased bases as the {incompatibility}, our bound will become extremely tight for the
three-qubit  {$\emph{X}$-state system}, completely
coinciding with the entropy-based uncertainty, and can restore Renes ${ {\emph{et al.}}}$'s bound with respect to arbitrary tripartite pure states.
In addition, by  applying our lower bound, one can attain the tighter bound of quantum secret key rate, which is of basic importance to enhance the security of quantum key
distribution protocols.
}

\end{abstract}

\maketitle

\section{Introduction}
Uncertainty principle proposed by Heisenberg is one of the pivotal cores in the area of quantum mechanics and exhibits
basic and clear {differences} distinguishing from its classical counterpart \cite{wh}. The uncertainty principle specifically
sets a lower bound for estimation of the measurement outcomes for two arbitrary incompatible observables on a quantum
system. Kennard \cite{ehk} and Robertson \cite{hpr} formulated the uncertainty principle in terms of a standard
deviation  $\Delta X \Delta Z \ge | {\langle \psi  |[ X ,\ Z ]| \psi  \rangle } |/2$
in regard to a pair of incompatible observables  $X$ and $Z$  for the systemic state
$\left| \psi  \right\rangle $.
Note that, it can be viewed that the lower bound of the relation is not an optimal prediction result,
because the bound is state-dependent, resulting in trivial result if the system is prepared in one of the eigenstates of  $X$ or $Z$.
Afterwards, there have been some efforts made to reform this relation
uncertainty and generalize the case of multi-observable \cite{lm,jll,qcs1,yx,aaa,rs,qcs2,WM2,cmn}. In 1983,
Deutsch took entropy measure into account depicting the uncertainty principle,
and conjectured the well-known form of entropic uncertainty relation (EUR) \cite{dd}.
 Further, Kraus improved Deutsch's uncertainty relation \cite{kk}, and later Maassen and Uffink proved the improvement \cite{hm}
\begin{align}
H\left( X \right) + H\left( Z \right) \ge  - {\log _2}c\left( {X,Z} \right) \equiv {q_{MU}},
\label{Eq.1}
\end{align}
where
$H\left(\tau \right) =  - \sum\nolimits_i {{p_i}{{\log }_2}{p_i}} $ is  the Shannon entropy of the measured
observables
$\tau  \in \left( {X,{\rm{ }}Z} \right)$ and
${p_i} = \left\langle {\Phi _i^\tau } \right|\rho \left| {\Phi _i^\tau } \right\rangle $ is the probability of
obtaining the $i$-th outcome for a measurement $\tau$.
$c\left( {X,Z} \right) \equiv {\max _{j,k}}{\left| {\langle \psi _j^X|\varphi _k^Z\rangle } \right|^2}$
 is maximal overlap,
and $\left| {\psi _j^X} \right\rangle $  and
$\left| {\varphi _k^Z} \right\rangle $ correspond to the eigenvectors of $X$ and $Z$,  respectively.
Since
$c\left( {X,Z} \right)$  is relevant to the two observables themselves, it directly shows that the lower bound of EUR is state-independent.
Compared with the  standard deviation, the EUR can enable us to better predict the measured uncertainty.

 The original conjecture of quantum cryptography \cite{sw} was inspired by the uncertainty relation.
 Yet, it is overlooked that the eavesdropper possibly has {the} entanglement  \cite{rh} with the measured
 system. Therefore, it cannot precisely prove the security of quantum cryptography by above-mentioned
 uncertainty relations. In  2010, {{for a bipartite system consisting of particles $A$ and $B$, treating $B$ as quantum memory,}}   Berta  {\it {\it et al.}}  \cite{mb}
 filled such a blank and generalized  EUR, i.e., quantum-memory-assisted entropic uncertainty relation (QMA-EUR), which reads
\begin{align}
S\left( {X|B} \right) + S\left( {Z|B} \right) \ge S\left( {A|B} \right) - {\log _2}c\left( {X,Z} \right),
\label{Eq.2}
\end{align}
where
$S\left( {X|B} \right){\rm{ = }}S\left( {{\rho _{XB}}} \right) - S\left( {{\rho _B}} \right)$
  denotes the conditional von Neumann entropy \cite{man}
of post-measurement state with
${\rho _{XB}} = $
$\sum\nolimits_i {\left( {{{\left| {\psi _i^X} \right\rangle }_A}\left\langle {\psi _i^X} \right| \otimes {\mathds{1}_B}} \right){\rho _{AB}}\left( {{{\left| {\psi _i^X} \right\rangle }_A}\left\langle {\psi _i^X} \right| \otimes {\mathds{1}_B}} \right)}$, likewise for ${\rho _{ZB}}$, and
${\mathds{1}_B}$ is an identical operator in the Hilbert space of $B$.
 And
$S\left( {A|B} \right){\rm{ = }}S\left( {{\rho _{AB}}} \right) - S\left( {{\rho _B}} \right)$
represents the conditional von Neumann entropy of systemic density operator
${\rho _{AB}}$ with
$S\left( {{\rho _{AB}}} \right) =  - {\rm{t}}{{\rm{r}}}\left( {{\rho _{AB}}{{\log }_2}{\rho _{AB}}} \right)$ and
${\rho _B} = {\rm{t}}{{\rm{r}}_A}\left( {{\rho _{AB}}} \right)$. Following this
novel inequality, several interesting results will be manifested: (i) if the measured particle $A$ is entangled with the memory particle $B$,
the conditional von Neumann entropy $S\left( {A|B} \right)$ can be negative, Bob's uncertainty for Alice's measured outcomes will be reduced; (ii)
when particle $A$   is maximally entangled with particle $B$, one can obtain
$S\left( {A|B} \right){\rm{ = }} - {\log _2}d$ with the dimension $d$ of the measured particle and this will lead to a  zero-valued bound, reflecting that
Bob can perfectly
predict the Alice's measured outcomes of both observables $X$  and $Z$; (iii) when the quantum memory is absent, Eq. (\ref{Eq.2}) can be reduced to
\begin{align}
H\left( X \right) + H\left( Z \right) \ge S\left( {\rho ^A} \right) - {\log _2}c\left( {X,Z} \right),
\label{Eq.3}
\end{align}
which offers a tighter bound in comparison with Maassen and Uffink's result, because of
$S\left( {\rho ^A} \right) \ge {\rm{0}}$ holds.

Actually, there have existed some promising {{improvement}} and generalizations related to QMA-EUR expressed by Eq. (\ref{Eq.2}) \cite{MAB,SW,AKP,TPSM, HML,TPP,LM,PJC,LR1,SZ,LR2,JZ,SL,FA,JLH}.
To be explicit, Pati  {\it et al.} \cite{AKP}
derived the uncertainty relation with a tighter lower bound including the classical correlation and
the quantum correlation quantified by quantum discord. And Pramanik {\it et al.} \cite{TPSM} reported a new form of uncertainty relation via extractable classical information.
Coles and Piani \cite{PJC} presented a strong bound via considering the second largest value of the overlap $c(X,Z)$.
Later on, Adabi {\it et al.} \cite{FA} presented the uncertainty lower bound by adding a term about  mutual information
and Holevo quantity. More recently, Huang  {\it et al.}  \cite{JLH} proposed a Holevo bound of QMA-EUR, which
yields an interesting result that the difference  between the entropic uncertainties and the new lower bound is
always a constant value.
So far, from an experimental viewpoint,  ongoing progress has been made by some groups \cite{Rp,Cfl,wcm11,zxc,wml,hyw11,zxc11,wcxy1}.

Typically, Renes and Boileau \cite{JJR} {{put forward}} the tripartite uncertainty relation and generalized it into arbitrary two measurements as
\begin{align}
S\left( {X|B} \right) + S\left( {Z|C} \right) \ge {q_{MU}},
\label{Eq.4}
\end{align}
where
{{$S\left( {X |B} \right){\rm{ = }}S\left( {{\rho _{X B}}} \right) - S\left( {{\rho _B}} \right)$ and
$S( {Z |C} ){\rm{ = }}S( {{\rho _{Z C}}} ) - S( {{\rho _C}} )$
  denote the conditional von Neumann entropy of post-measurement states,}} and
$q_{MU}$ is same as in Eq. (\ref{Eq.1}). Technically, there is a trade-off that is quantified by the
complementarity of the measurements. And this relation can be interpreted by a guessing game, so-called monogamy game.
Explicitly, there are three participants, Alice, Bob and Charlie in this game. Preparing a tripartite state
${\rho _{ABC}}$,
particle $A$ is sent to Alice,  $B$  to Bob and   $C$ to Charlie.
After receiving $A$, Alice randomly chooses a measurement ($X$ or $Z$) and obtains the corresponding measure outcome $\epsilon$.
Then Alice informs Bob and Charlie of her measurement choice. Bob and Charlie will win this game if and only if both predict $\epsilon$.
By utilizing the monogamy of entanglement, Eq. (\ref{Eq.4}) shows the uncertainty via the game: if Bob correctly produces a guess
in case that Alice measured $X$ on $A$, as a result, Charlie cannot produce a good guess in case that Alice measured $Z$ on $A$,  and vice versa.
With respect to  given observables $X$ and $Z$, the bound $q_{MU}$  will become a constant, which will be independent of the characteristic of the system to be probed.
In principle, the bound should be associated with the system.
Although there have existed some
improvement {{regarding}}  entropic uncertainty relation, these explorations are confined to bipartite systems.
Till now,  there have been few improvement of tripartite quantum-memory uncertainty relations.
Here we put forward a tighter bound of tripartite uncertainty relation with quantum memory,
which resorts to mutual information and Holevo quantity, and thus is applicable to
a lower bound of the tighter bound of quantum secret key rate to enhance the security of quantum key
distribution protocols.

The outline of this article is organized as follows. In Sec. II, we derive a new tripartite uncertainty relation for arbitrary tripartite state,
and manifest that the new bound is tighter than the
previous bound. In Sec. III, we investigate our presented lower bound for several examples (generalized GHZ state, generalized $W$ state, Werner-type state and random three-qubit
states) to show its performance. In Sec. IV, the application of our result on quantum secret key rate is discussed. Lastly, the concise
conclusions and discussions are given  in Sec. V.

\section{Improved tripartite entropic uncertainty relation}
Here, we present a brand-new and tighter lower bound for tripartite uncertainty relation with quantum memory, as an improvement of
the existed uncertainty relation expressed as Eq. (\ref{Eq.4}). \vskip 0.2cm

{\bf Theorem.} By considering mutual information and Holevo quantity, a new tripartite uncertainty relation can be obtained as

\begin{align}
S\left( {X|B} \right) + S\left( {Z|C} \right) \ge {q_{MU}} + \max \left\{ {0,\Delta } \right\}
\label{Eq.5}
\end{align}
with
\begin{align}
\Delta  = &{q_{MU}} + 2S\left( {\rho ^A} \right) - \left[ {{\cal I}\left( {A:B} \right) + {\cal I}\left( {A:C} \right)} \right]\nonumber \\
 & + \left[ {{\cal I}\left( {Z:B} \right) + {\cal I}\left( {X:C} \right)} \right] - H\left( X \right) - H\left( Z \right),
\label{Eq.6}
\end{align}
where
${\cal I}\left( {A;B} \right) = S\left( {{\rho ^A}} \right) + S\left( {{\rho ^B}} \right) - S\left( {{\rho ^{AB}}} \right)$ is
mutual information. The Holevo quantity
${\cal I}\left( {X;B} \right) = S\left( {{\rho ^B}} \right) - \sum\limits_i {{p_i}S\left( {\rho _i^B} \right)} $ \cite{FA}
denotes the upper bound of accessible information of Bob for Alice's measurement outcome.
When Alice performs measurement $X$ on particle $A$, and obtain the $i$-th measurement outcome with
probability
${p_i} = {{\mathop{\rm tr}\nolimits} _{AB}}\left( {\Pi _i^A{\rho ^{AB}}\Pi _i^A} \right)$, and particle $B$ corresponds to the state
$\rho _i^B = \frac{{{\mathop{\rm Tr}\nolimits} \left( {\Pi _i^A{\rho ^{AB}}\Pi _i^A} \right)}}{{{p_i}}}$.
{{$H\left( X \right)$ is the Shannon entropy  by performing  the measurement $X$ on subsystem $A$ (likewise for $H\left( Z \right)$).}}
The dynamics of entropic uncertainty  can be reflected by the evolution of tighter lower bound. This is a remarkable result that the new lower bound can capture the characteristic how the entropic uncertainty would behave.

{\bf Proof.}  Based on QMA-EUR in Eq. (\ref{Eq.2}), with regard to the subsystems $B$ and  $C$, we have
\begin{align}
S\left( {X|B} \right) + S\left( {Z|B} \right) \ge S\left( {A|B} \right) + {q_{MU}},
\label{Eq.A1}
\end{align}
\begin{align}
S\left( {X|C} \right) + S\left( {Z|C} \right) \ge S\left( {A|C} \right) + {q_{MU}}.
\label{Eq.A2}
\end{align}
By combining Eqs. (\ref{Eq.A1}) and (\ref{Eq.A2}), a new inequality can be derived by
\begin{align}
S\left( {X|B} \right) + S\left( {Z|C} \right) &\ge 2{q_{MU}} + S\left( {A|B} \right) + S\left( {A|C} \right) \nonumber\\
&{\rm{                                  }} - S\left( {Z|B} \right) - S\left( {X|C} \right).
\label{Eq.A3}
\end{align}
Making use of the relations  $S\left( {\rho ^A} \right){\rm{ = }}S\left( {A|B} \right) + {\cal I}\left( {A;B} \right)$,  $S\left( {\rho ^A} \right){\rm{ = }}S\left( {A|C} \right) + {\cal I}\left( {A;C} \right)$,
$H\left( Z \right){\rm{ = }}S\left( {Z|B} \right) + {\cal I}\left( {Z;B} \right)$ and
$H\left( X \right) = S\left( {X|C} \right) + {\cal I}\left( {X;C} \right)$,
and resorting to Eqs. (\ref{Eq.4}) and (\ref{Eq.A3}),
the tripartite quantum memory uncertainty relation can be reformulated into the desired outcome, i.e., Eq. (\ref{Eq.5}).

Noting that there are some special cases that $\Delta $ can be reduced. One is that if $X$ and $Z$ are
complementary observables and subsystem $A$ is a maximally mixed state, we have $H\left( X \right) + H\left( Z \right) {\rm{ = }} {q_{MU}} + S\left( {\rho ^A} \right)$, such as GHZ state.
And, another case is that the observables are Pauli measurements $\sigma_x$ and $\sigma_z$, and the subsystem  $A$ is an incoherent state, this equality mentioned above holds, such as generalized GHZ state, generalized $W$ state Werner-type state. Hence, in both cases,
$\Delta\equiv S\left( {\rho ^A} \right) - \left[ {{\cal I}\left( {A:B} \right) + {\cal I}\left( {A:C} \right)} \right] + \left[ {{\cal I}\left( {Z:B} \right) + {\cal I}\left( {X:C} \right)} \right]$, which
obtains a reduced form compared with the previous.
\vskip 0.2cm

{\bf Corollary 1.} If the choosing observables  $X$  and  $Z$  are mutual-unbiased bases
${\sigma_x}$ and ${\sigma_z}$ on ${\cal {H}}_A$, and the prepared state is with form of an arbitrary three-qubit $X$-structure state,
our lower bound in Eq. (\ref{Eq.5}) is extremely tight, because our bound
${q_{MU}} + \max \left\{ {0,\Delta } \right\}$ will coincide
with the sum of Bob's entropic uncertainty and Charlie's entropic uncertainty
$S\left( {X|B} \right) + S\left( {Z|C} \right)$.
According to lower bound, we can precisely predict the sum of Bob's uncertainty and Charlie's uncertainty in such a case.

{\bf Proof.} For a class of three-qubit $X$-structure states,  these states can be denoted as
\begin{align}
{\rho ^X}{\rm{ = }}\left( {\begin{array}{*{20}{c}}
{{\rho _{11}}}&0&0&0&0&0&0&{{\rho _{18}}} \\
0&{{\rho _{22}}}&0&0&0&0&{{\rho _{27}}}&0 \\
0&0&{{\rho _{33}}}&0&0&{{\rho _{36}}}&0&0 \\
0&0&0&{{\rho _{44}}}&{{\rho _{45}}}&0&0&0 \\
0&0&0&{{\rho _{45}}}&{{\rho _{55}}}&0&0&0 \\
0&0&{{\rho _{36}}}&0&0&{{\rho _{66}}}&0&0 \\
0&{{\rho _{27}}}&0&0&0&0&{{\rho _{77}}}&0 \\
{{\rho _{18}}}&0&0&0&0&0&0&{{\rho _{88}}}
\end{array}} \right)
\label{Eq.B1}
\end{align}
in orthogonal basis
$\{| {000} \rangle, | {001} \rangle, | {010} \rangle,  | {011}\rangle, | {100} \rangle, | {101} \rangle,$
$ | {110} \rangle, | {111} \rangle \}$.
${\rho _{ij}}\left( {i,j = 1,2,3,4,5,6,7,8} \right)$ are all real parameters, and it satisfies the normalized
condition $\sum_{i=1}^8\rho_{ii}=1$. Choosing two Pauli measurements
${\sigma _x}$  and ${\sigma _z}$  performed on the particle  $A$,
we can derive the analytical solution of the sum of Bob's uncertainty and Charlie's
uncertainty  ${U_{L}} = S\left( {X|B} \right) + S\left( {Z|C} \right)$ (the left-hand side of Eq. (\ref{Eq.5})) about Alice's measurement outcome
and the uncertainty lower bound ${U_R} = {q_{MU}} + \max \left\{ {0,\Delta } \right\}$ (the right-hand side of Eq. (\ref{Eq.5})) under the normalization condition as following
\begin{align}
{U_L} = {U_R} &= 1 - {S_{bin}}\left( {{\rho _{11}} + {\rho _{33}} + {\rho _{55}} + {\rho _{77}}} \right) \nonumber \\
&{\rm{          }} - \left( {{\rho _{11}} + {\rho _{33}}} \right){\log _2}\left( {{\rho _{11}} + {\rho _{33}}} \right) \nonumber \\
&{\rm{          }} - \left( {{\rho _{22}} + {\rho _{44}}} \right){\log _2}\left( {{\rho _{22}} + {\rho _{44}}} \right) \nonumber \\
&{\rm{          }} - \left( {{\rho _{55}} + {\rho _{77}}} \right){\log _2}\left( {{\rho _{55}} + {\rho _{77}}} \right) \nonumber \\
&{\rm{          }} - \left( {{\rho _{66}} + {\rho _{88}}} \right){\log _2}\left( {{\rho _{66}} + {\rho _{88}}} \right),
\label{Eq.B2}
\end{align}
where
${S_{bin}}\left( \Upsilon  \right) =  - \Upsilon {\log _2}\Upsilon  - \left( {1 - \Upsilon } \right){\log _2}\left( {1 - \Upsilon } \right)$
represents the binary entropy. Therefore, our lower bound is always equal to  the sum of Bob's uncertainty and Charlie's uncertainty
in the current architecture.\vskip 0.2cm

{\bf Corollary 2.} If the prepared state is a tripartite pure state, owing to that $\Delta $  is always less than or equal
to zero, our lower bound will recover Renes {\it et al.}'s result.

{\bf Proof.} For an any tripartite pure state, we have that the conditional entropies satisfy
\begin{align}
S\left( {A|B} \right) + S\left( {A|C} \right) = 0.
\label{122}
\end{align}
From Eq. (\ref{Eq.4}), one easily obtains
\begin{align}
{q_{MU}} - S\left( {Z|B} \right) - S\left( {X|C} \right) \le 0.\label{123}
\end{align}
By linking Eqs. (\ref{Eq.5}) and (\ref{Eq.A3}) with Eqs. (\ref{122}) and (\ref{123}), it is obtained that $\Delta $ always less than or equal to zero,
and our lower bound will recover   Renes  {\it et al.}'s lower bound. Moreover,
we reveal the relationship between our lower and the purity of the systemic state (the  purity
$P({\rho _{ABC}}) = {\mathop{\rm Tr}\nolimits} \left[ {{\rho ^2_{ABC}}} \right]$) by means of the approach of random states.
That is our lower bound decreases with the increasing purity, and  Renes  {\it et al.}'s lower bound will be restored once
the purity maximizes (i.e., the case of pure states).

\section{Examples}
Considering a pair of incompatible observables such that perfect knowledge about observable $X$ implies complete
ignorance about observable $Z$, the observables are called unbiased or mutually unbiased. For any finite-dimensional
space, there are many pairs of orthogonal bases that satisfy this property. On the another hand, if two orthogonal
bases $X$ and $Z$ are mutually unbiased bases (MUBs), they must satisfy the condition
${\left|\langle {\psi _j^X|\varphi _k^Z\rangle } \right|^2} := {1}/{d}\ \left( {\forall {\rm{ }}j,k} \right)$.
For example, if the measured particle is a qubit, Pauli measurements
${\sigma _x}$, ${\sigma _y}$ and  ${\sigma _z}$ can be chosen as the incompatible observables and MUBs.
Generally, spin-1/2 Pauli matrices can be written as
${\sigma _x} = \left| 0 \right\rangle \left\langle 1 \right| + \left| 1 \right\rangle \left\langle 0 \right|$,
${\sigma _y} =  - i\left| 0 \right\rangle \left\langle 1 \right| + i\left| 1 \right\rangle \left\langle 0 \right|$
and ${\sigma _z} = \left| 0 \right\rangle \left\langle 0 \right| - \left| 1 \right\rangle \left\langle 1 \right|$, which
form a set of three MUBs. Through
calculating the eigenvectors of Pauli matrices, the maximal overlap
$c\left( {X,Z} \right) \equiv {\max _{j,k}}{\left| {\langle \psi _j^X|\varphi _k^Z\rangle } \right|^2}$ is always $\frac{1}{2}$. Thereupon, the incompatible term is
$- {\log _2}c\left( {X,Z} \right) = 1$. As an illustration, we herein choose Pauli measurements  ${\sigma _x}=X$ and  ${\sigma _z}=Z$   as the incompatibility,
and discuss our result in different scenarios.

%

\subsection{{Generalized GHZ state}}
First of all, let us consider a typical pure tripartite state, generalized GHZ state,
which can be written as in the Schmidt basis
\begin{align}
{\left| \psi  \right\rangle _{\rm {GHZ}}} = \cos \beta \left| {000} \right\rangle  + \sin \beta \left| {111} \right\rangle
\label{Eq.01}
\end{align}
with $\beta\in[ 0,\pi/2 ]$.
After performing two incompatible observables on particle $A$,  the subsystems $B$ and $C$ will become pure states,
$S\left( {\rho_B} \right) = S\left( {\rho_C} \right)$ and $S\left( {\rho_{ZC}} \right)=S\left( {\rho_C} \right)$. As a result, we obtain $S\left( {Z|C} \right)=0$,
and the entropic uncertainty only relies on the conditional von Neumann entropy $S\left( {X|B} \right)$.
Interestingly,  the conditional von Neumann entropy $S\left( {X|B} \right)=1$ is invariable.
Besides, we can compute the improved bound is valued-one as well, { {which verifies the result in Corollary 2.
Choosing Pauli measurements as the incompatibility, Renes {\it et al.}'s lower bound (the right-hand side of Eq. (\ref{Eq.4})) is
$q_{MU}=- {\log _2}c\left( {X,Z} \right) = 1$. Therefore, the Numerical  solution of the sum of Bob's uncertainty and Charlie's
uncertainty  ${U_{L}}$ (the left-hand side of Eq. (\ref{Eq.5})) about Alice's measurement outcome, the uncertainty's lower bound ${U_R}$ (the right-hand side of Eq. (\ref{Eq.5})) and Renes {\it et al.}'s lower bound are all equal to valued-one.}}

\begin{figure}
\centering
\includegraphics[width=7.5cm]{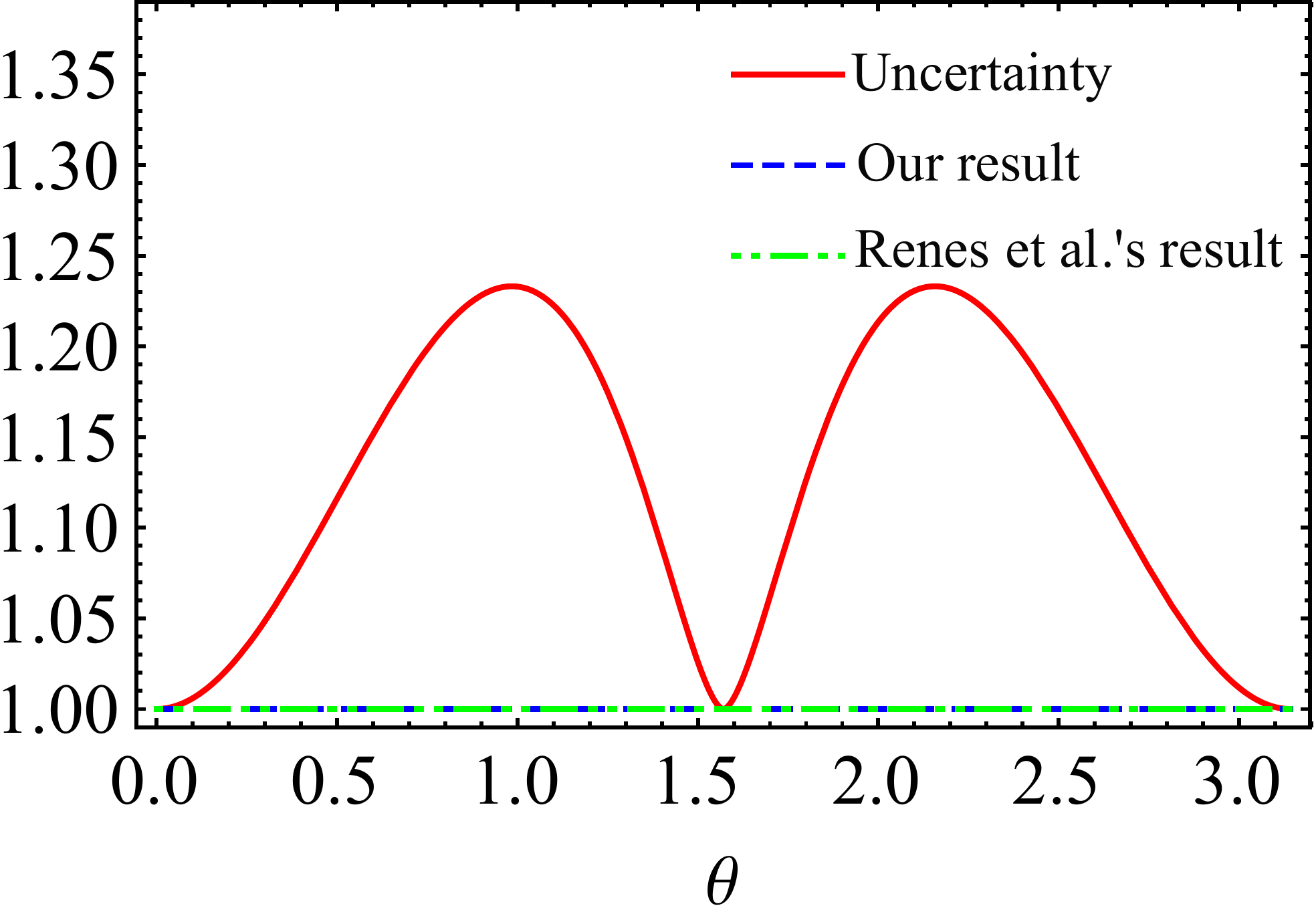}
\caption{(Color online)  Uncertainty and the lower bounds versus the state's parameter
$\theta $. The red solid line  represents entropic uncertainty (the left hand-side of Eq. (\ref{Eq.5})),
the blue dashed line shows our result (the right-hand side of Eq. (\ref{Eq.5}))
and the green dashed line shows Renes {\it et al.}'s result (the right hand-side of Eq. (\ref{Eq.4})).}
\label{f1}
\end{figure}

\begin{figure}
\centering
\includegraphics[width=7.5cm]{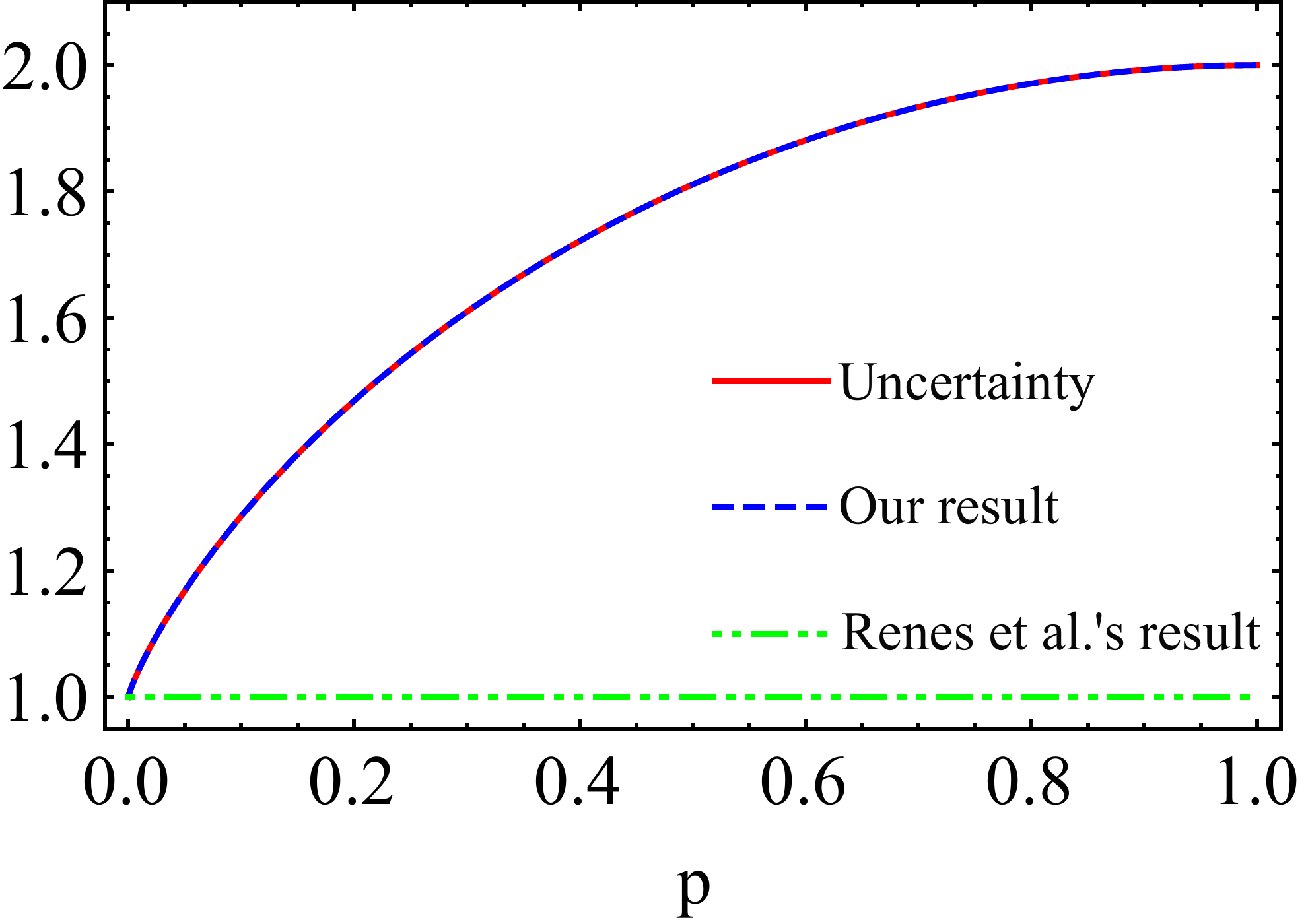}
\caption{(Color online)  Uncertainty and the lower bounds versus the state's parameter
$\theta $. The red solid line  represents entropic uncertainty (the left hand-side of Eq. (\ref{Eq.5})),
the blue dashed line shows our result (the right-hand side of Eq. (\ref{Eq.5}))
and the green dashed line shows Renes {\it et al.}'s result (the right hand-side of Eq. (\ref{Eq.4})).}
\label{f2}
\end{figure}

\subsection{{Generalized $W$ state}}
 As another example, we take into account the generalized $W$ state in the Hilbert space spanned by
$\left\{ {\left| 0 \right\rangle ,\left| 1 \right\rangle } \right\}$  as
\begin{align}
{\left| \psi  \right\rangle _{\mathop{\rm W}\nolimits} } = \cos \theta  \left| {001} \right\rangle  + \sin\theta\cos\alpha\left| {010} \right\rangle  + \sin \theta \sin \alpha \left| {100} \right\rangle.
\label{Eq.20}
\end{align}
As can be seen from Fig. \ref{f1}, our lower bound remain a constant in all intervals related to parameter $\theta$.
Owning to $S\left( {A|B} \right) + S\left( {A|C} \right) = 0$ and ${q_{MU}} - S\left( {Z|B} \right) - S\left( {X|C} \right) \le 0$, our lower bound
is identical with the lower bound presented
by Renes {\it et al.}, which also proofs our statement made before.

\subsection{{Werner-type state}}
Supposing that Alice, Bob and Charlie initially share a Werner-type state under
the basis $\left\{ {\left| 0 \right\rangle ,\left| 1 \right\rangle } \right\}$  as
\begin{align}
{\left| \psi \right\rangle}_{\rm Werner} = \left( {1 - p} \right)\left| \Phi \right\rangle \left\langle \Phi \right| + \frac{p}{8}{\mathds{1}_{8\times8}},
\label{Eq.20}
\end{align}
where the GHZ state
$\left| \Phi  \right\rangle=\frac{{\sqrt 2 }}{2}\left( {\left| {000} \right\rangle  + \left| {111} \right\rangle } \right)$,   ${\mathds{1}_{8\times8}}$
stands for an identity $8\times 8$ matrix and  $0 \le p \le 1$.
In Fig. \ref{f2}, considering Werner-type state as a specific $X$ state and two specific incompatible observables,  the entropic uncertainty and
our lower always are synchronized all the time. This reflects that our bound not only  depends on the observables, but on the initial state.
It shows that our lower bound is tighter than previous lower bound for all $p$, which essentially is in agreement with Corollary 1,  because $\Delta\geq0$ holds in the case. \vskip 3cm

\subsection{{Random three-qubit states}}

\begin{figure}
\begin{minipage}{0.45\textwidth}
\centering
\subfigure{\includegraphics[width=7.6cm]{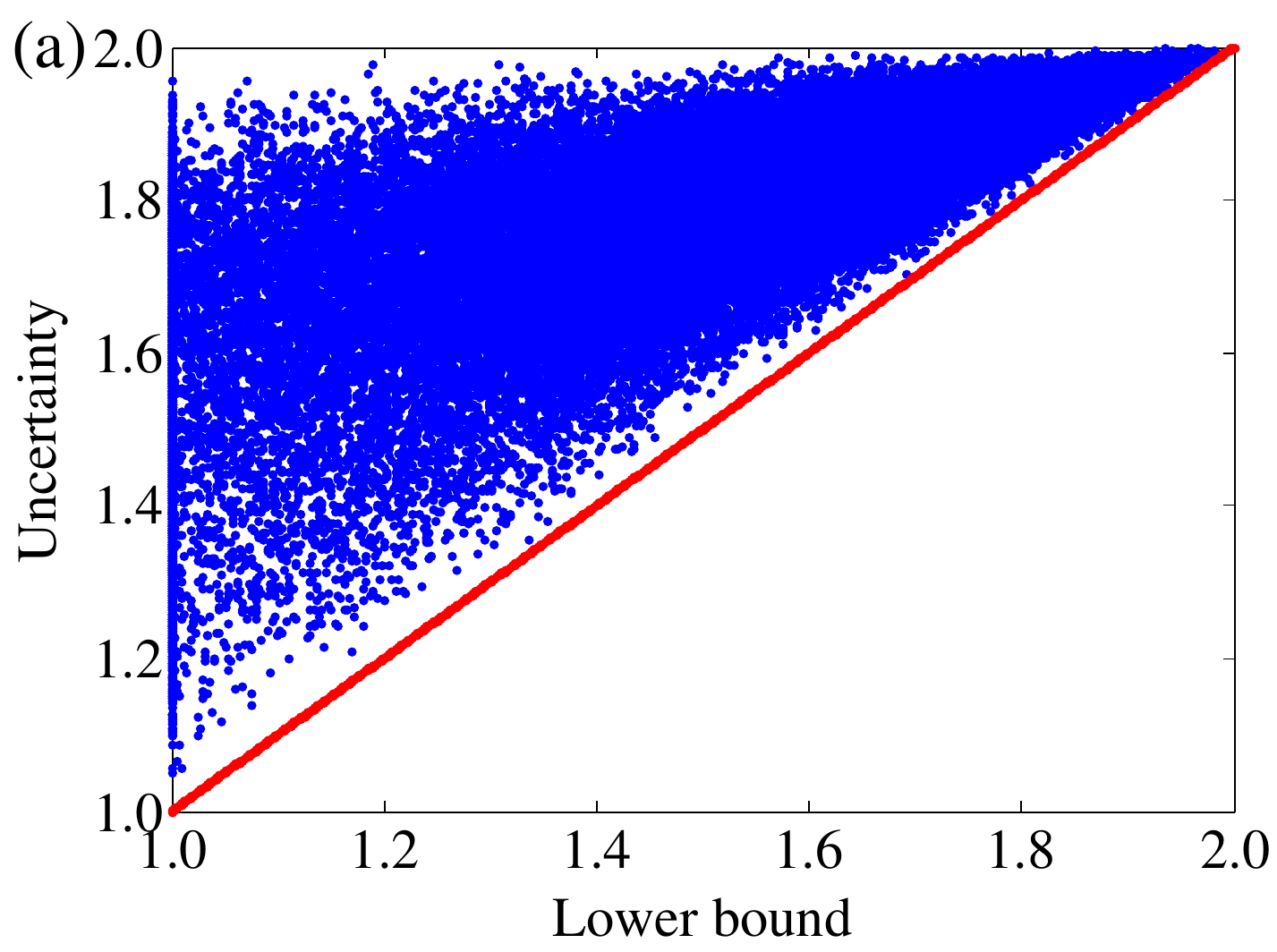}}
\subfigure{\includegraphics[width=7.5cm]{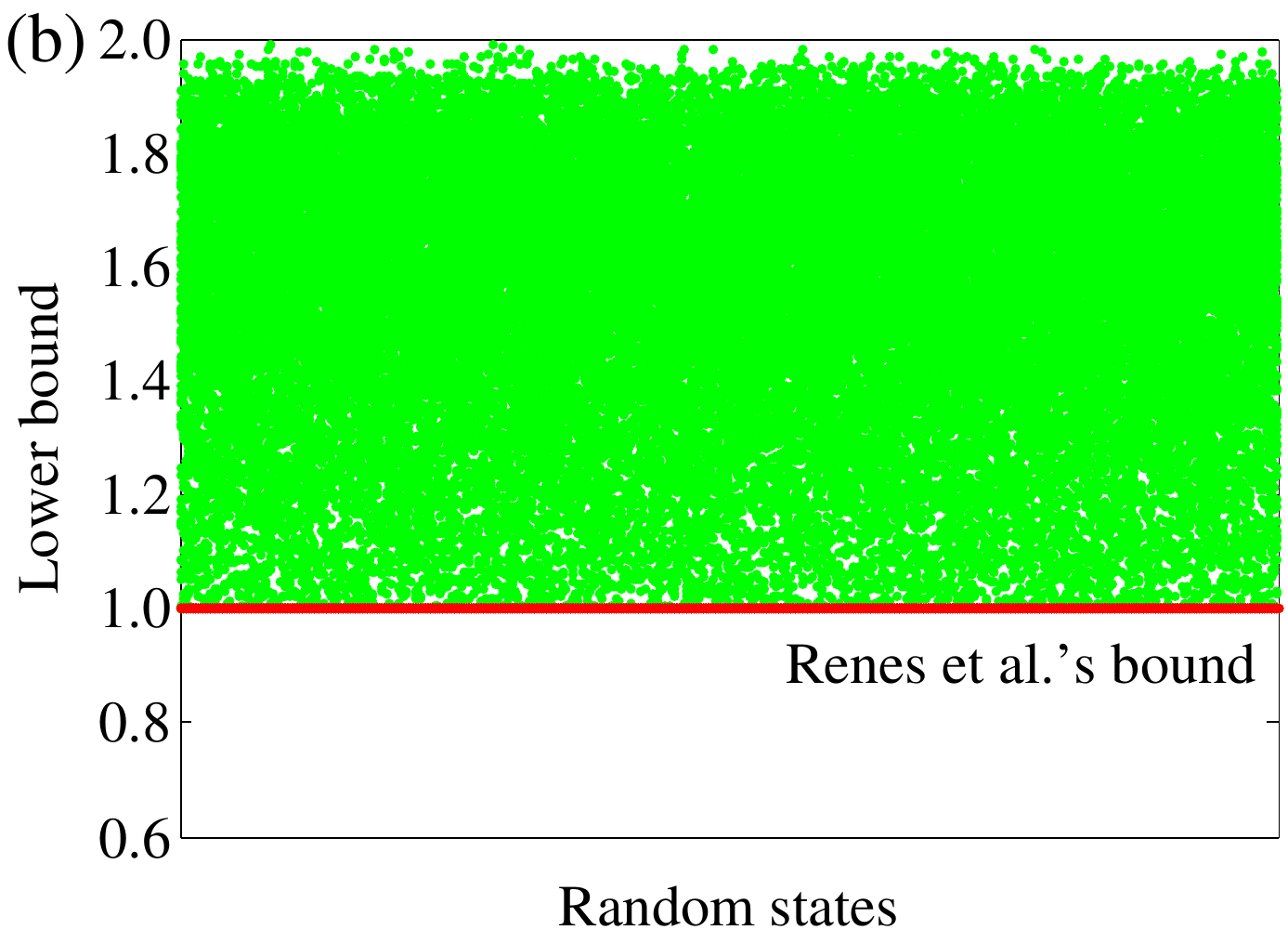}}
\end{minipage}\hfill
\caption{(Color online) {{(a) Uncertainty $U_L$ versus our lower bound $U_R$ for
${10^5}$ randomly generated three-qubit states.  X-axis represents our lower bound and
Y-axis denotes the entropic uncertainty, respectively. The red line denotes the proportional function with the slope of
unity. (b) Comparison between our lower bound $U_R$ and Renes {\it et al.}'s lower bound for
${10^5}$ randomly generated three-qubit states.  X-axis stands for randomly generated three-qubit states,  and
Y-axis denotes the lower bound. The green scattering points are the amount of our
bound and the red line represents Renes {\it et al.}'s  bound.}}}
\label{f3}
\end{figure}

Now let us introduce the construction of random three-qubit states.
{{Here, we adopt an effective approach to generate random density matrices, which is different from the promising method by Haar measure \cite{Karol}.}}
Take the example of  generating an arbitrarily random three-qubit state,
an arbitrary three-qubit state $\rho $ can be decomposed by the state's eigenvalues
${\lambda _n}$  and normalized eigenvectors
$\left| {{\psi _n}} \right\rangle $, i.e.,
$\rho  = \sum\limits_{n = 1}^8 {{\lambda _n}\left| {{\psi _n}} \right\rangle \left\langle {{\psi _n}} \right|} $ with ($n \in \left\{ {1,2,3,4,5,6,7,8} \right\}$). The eigenvalue ${\lambda _n}$  corresponds to the probability that the state $\rho$
is in the pure state $\left| {{\psi _n}} \right\rangle $. And the state's normalized eigenvectors
$\left| {{\psi _n}} \right\rangle $  can establish an arbitrary unitary operation
$E = \left\{ {\left| {{\psi _1}} \right\rangle ,\left| {{\psi _2}} \right\rangle ,\left| {{\psi _3}} \right\rangle ,\left| {{\psi _4}} \right\rangle ,\left| {{\psi _5}} \right\rangle ,\left| {{\psi _6}} \right\rangle ,\left| {{\psi _7}} \right\rangle ,\left| {{\psi _8}} \right\rangle } \right\}$.
An arbitrary three-qubit state also can be composed by an arbitrary set of probabilities ${\lambda _n}$   and an arbitrary $E$.
Hence, we can construct an arbitrary three-qubit state by an arbitrary set of probabilities and an arbitrary unitary operation.
The random number function $f\left( {{x_1},{x_2}} \right)$ randomly generate a real number in a closed interval
$[ {{x_1},{x_2}} ]$. At start, we can generate eight random numbers by  this method
\begin{align}
{{\cal P}_1} &= f\left( {0,1} \right),\nonumber\\
{{\cal P}_2} &= f\left( {0,1} \right){{\cal P}_1},\nonumber\\
{{\cal P}_3} &= f\left( {0,1} \right){{\cal P}_2},\nonumber\\
{{\cal P}_4} &= f\left( {0,1} \right){{\cal P}_3},\nonumber\\
{{\cal P}_5} &= f\left( {0,1} \right){{\cal P}_4},\nonumber\\
{{\cal P}_6} &= f\left( {0,1} \right){{\cal P}_5},\nonumber\\
{{\cal P}_7} &= f\left( {0,1} \right){{\cal P}_6},\nonumber\\
{{\cal P}_8} &= f\left( {0,1} \right){{\cal P}_7}.
\label{Eq.A5}
\end{align}
Then a set random probabilities ${\lambda _n}$  ($n \in \left\{ {1,2,3,4,5,6,7,8} \right\}$)
are controlled by the random numbers ${\cal P}_m$ expressed as
\begin{align}
{\lambda _n} = \frac{{{{\cal P}_m}}}{{\sum\limits_{m = 1}^8 {{{\cal P}_m}} }}.
\label{Eq.A6}
\end{align}
From Eqs. (\ref{Eq.A5}) and (\ref{Eq.A6}), it is straightforward to get a set of random probabilities in  descending order.

For the random generation of unitary operation, we first randomly give an 8-order real matrix $T$ by the random number function $f\left( {{-1},{1}} \right)$ with the closed interval $\left[ { - 1,1} \right]$.
Using the real matrix  $T$, a random Hermitian matrix can be given as

\begin{align}
H = D + \left( {{U^ \top } + U} \right) + i\left( {{L^ \top } - L} \right),
\label{Eq.A7}
\end{align}
where
$U$, $L$ and $U$ denote diagonal, strictly lower- and strictly upper-triangular part of the real matrix $T$ respectively, and $U^\top$ stands for the transposition of the matrix $U$.
By the calculation, we can attain eight normalized
eigenvectors $\left| {{\psi _n}} \right\rangle $ of the Hermitian matrix $H$, which forms the random unitary operation $E$.
Thus, we can construct the random three-qubit states
by the expression $\rho  = \sum\limits_{n = 1}^8 {{\lambda _n}\left| {{\psi _n}} \right\rangle \left\langle {{\psi _n}} \right|} $.
As an illustration,
we take $10^5$ random states to depict the corresponding uncertainty and our bounds in Fig. \ref{f3}(a). It
is apparent to show that, $U_L\geq U_R$ (Eq. (\ref{Eq.5})) satisfies always.
By means of utilizing random states, we verify that our theorem presented here is hold.
{{Because Renes {\it et al.}'s lower bound is only dependent on the observable choices, it always remains a fixed value, it is apparent that our bound is greater than or equal to Renes {\it et al.}'s bound in Fig. \ref{f3}(b).}}

Additionally, following Fig. \ref{f4}, it can be easily found that the lower bound $U_R$  obtained by us gradually reduce
with the increasing systemic state purity $P$. For $P=\frac{1}{8}$  corresponding
to the purity of maximum mixed state, our lower bound will reach to the maximal  value  $U_R=2$.
When  $P=1$ corresponds to the three-qubit pure state, our lower bound will reduce to the minimal value
$U_R=1$.  In this case, our lower bound will restore Renes {\it et al.}'s lower bound.

\begin{figure}
\centering
\includegraphics[width=7.5cm]{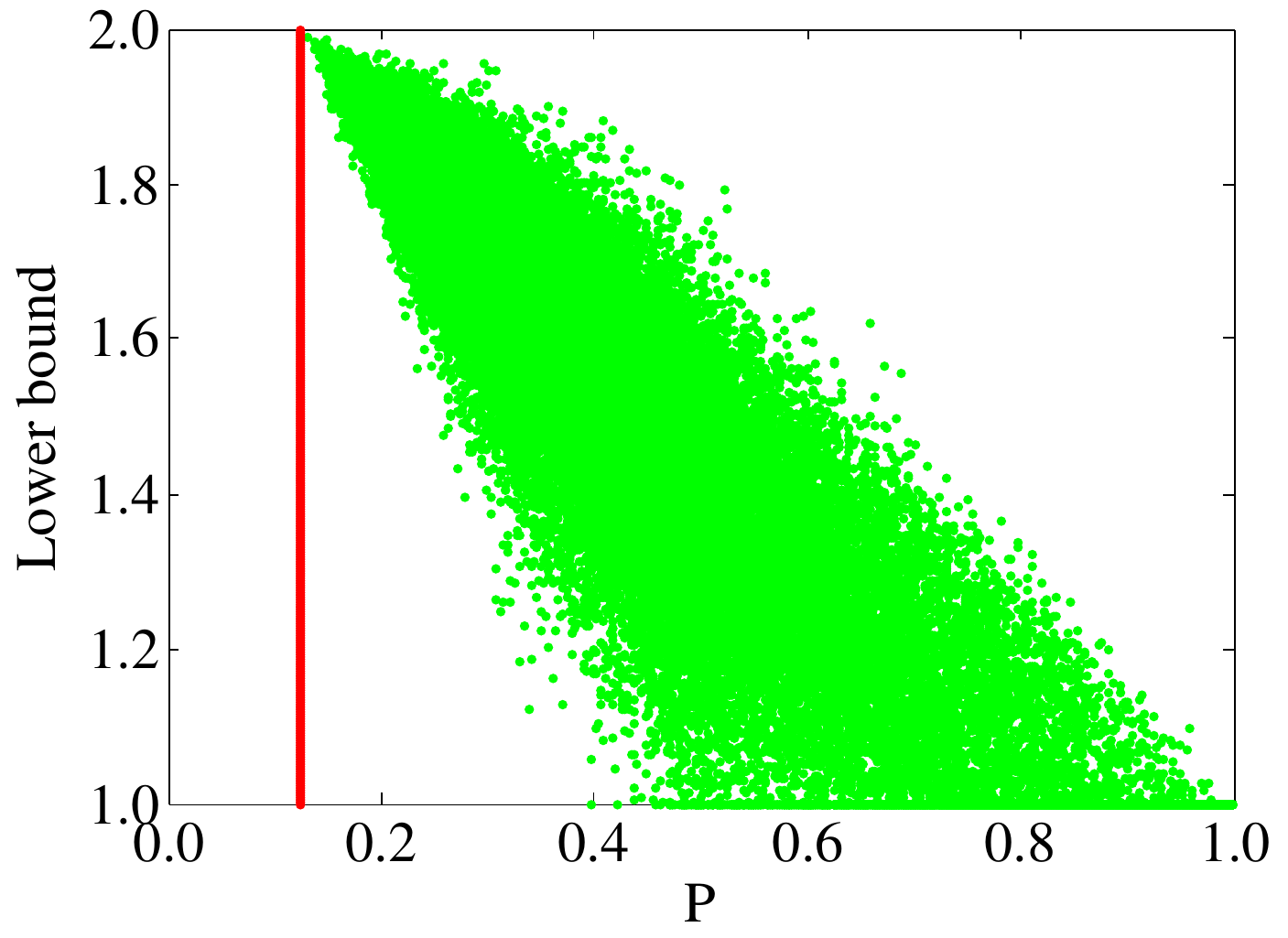}
\caption{(Color online) Our lower bound $U_R$ versus the systemic state purity with respect to
${10^5}$ randomly generated three-qubit states.  X-axis denotes the systemic state purity and  Y-axis represents our lower bound, respectively. The red line stands for
the case of ${P={\frac{1}{8}}}$.}
\label{f4}
\end{figure}

\section{Application}

Entropic uncertainty relation not only reflects the fundamental discrepancy between quantum mechanics and classical counterpart,
but also gives rise to  many potential applications in the course of quantum information processing, including
entanglement criterion \cite{Mhp,Yh}, quantum randomness \cite{Gv}, quantum steering \cite{aa1,aa2,aa3,ccjl}, quantum key distribution \cite{Id,Pjc,mb,JJR}, and so on.
Here we focus on the application of our finding on quantum key distribution. Specifically,
we derive quantum secret key rate lower bound based on the lower bound of tripartite uncertainty relation.
Technically, the entropic uncertainty relations can be applied to confirm the security of quantum key distribution protocols. To be specific,
the lower bound of tripartite uncertainty relation is closely associated with the quantum secret key (QSK) rate.
The key distribution protocol is that two honest part (Alice and Bob) share a key together by communicating over
a public channel, and the key is secret from any eavesdropping by the third part (Eve). Devetak and Winter \cite{Id}
reported that the amount of key $K$ that can be extracted by Alice and Bob is lower bounded by
\begin{align}
K \ge S\left( {Z|E} \right) - S\left( {Z|B} \right),
\label{Eq.22}
\end{align}
where Eve (eavesdropper) prepares a quantum state  $\rho_{ABE}$ and send particles $A$ and
$B$ to Alice and Bob respectively and retains $E$. Based on Renes {\it et al.}'s result in Eq. (\ref{Eq.4}) as
$S\left( {X|B} \right) + S\left( {Z|E} \right) \ge {q_{MU}}$,
we have
\begin{align}
K \ge {q_{MU}} - S\left( {X|B} \right) - S\left( {Z|B} \right).
\label{Eq.23}
\end{align}
According to the new lower bound of tripartite uncertainty relation in Eq. (\ref{Eq.5}),
the bound of quantum secret key rate can be rewritten as
\begin{align}
K' \ge {q_{MU}} + \max \left\{ {0,\Delta } \right\} - S\left( {X|B} \right) - S\left( {Z|B} \right).
\label{Eq.24}
\end{align}
Since the additional term
$\max \left\{ {0,\Delta } \right\}$ is greater than or equal to zero all along, we declare that the QSK rate
obtained by us is tighter than the result obtained by Berta {\it et al.} \cite{mb}.
It is well known that any measurements cannot reduce {{the systemic}} entropy. Thus, the bound of quantum secret key rate can be derived as
 \begin{align}
K' \ge {q_{MU}} + \max \left\{ {0,\Delta } \right\} - S\left( {X|X'} \right) - S\left( {Z|Z'} \right).
\label{Eq.25}
\end{align}
When conjugate observables are applied to qubits and assumed symmetric, we have  $S\left( {X|X'} \right) =S\left( {Z|Z'} \right)$.
Although the argument applies only to collective attacks, it can be extended to arbitrary attacks via the post-selection technique \cite{mcrk}.
The advantage of security argument is that Alice and Bob only need to upper bound the additional term
$\max \left\{ {0,\Delta } \right\}$ and the entropies $S\left( {X|X'} \right)$ and $S\left( {Z|Z'} \right)$,
 which it can improve the performance of the actual quantum key distribution protocols.
 The statistics required to estimate states are critical for the security of protocols \cite{rrvs}.
We can analyze the security of quantum
key distribution protocols  by assuming that the
eavesdropper creates a quantum state $\rho_{ABE}$, and sends particles $A$ and
$B$ to Alice and Bob, respectively. Although in this case, a security proof certainty means security
when Alice and Bob distribute the states themselves \cite{mb}.
In order to generate their key,
Alice and Bob  randomly choose the  measurements and measure the states. $X$ and $Z$ are Alice's choosing
measurements, and $X'$ and $Z'$ are Bob's measurements.
To ensure that the same key can be generated,  Alice and Bob inform each other of their measurement choices.
In the worst case, the communication between Alice and Bob is completely
overheard by an eavesdropper Eve who try to get the key.
Herein, based on the Devetak-Winter formula \cite{Id}, the referring communication
corresponds to the error correction and privacy amplification steps in the quantum key distribution protocol.
Berta {\it et al.} declared that if the measurement outcomes are sufficiently correlated, the quantity of key $K$
will be positive which means key can be asymptotically generated \cite{mb}. Compared with Berta {\it et al.}'s
result $K$, our result
$K'$  adds a term $\max \left\{ {0,\Delta } \right\}$, which is stronger than the previous.
Thereby, even in this case, if the measurement outcomes are sufficiently correlated,
the quantum secret key $K'$ will be positive, leading to that Alice and Bob can still generate a secure key.

\section{Discussions and conclusions}
We have derived a lower bound of tripartite uncertainty relation with quantum memory by adding an
additional term related with mutual information and Holevo quantity. It has been demonstrated that
our lower bound outperform the previous bound to some extent. We prove that our lower bound will completely coincide
with the total preparation uncertainty, with regard to ${\sigma _x}$ and ${\sigma _z}$ as the incompatibility in the framework
of an arbitrary three-qubit $X$-structure state. For an arbitrary tripartite pure state, since
the quantity $\Delta$  is less than or equal
to zero at all, our lower bound can recover Renes {\it et al.}'s lower bound.
As illustrations, we specifically take into account the tripartite
uncertainty relation for generalized GHZ state, generalized $W$ state, Werner-type state and random three-qubit states.
By the analytical calculation, it testifies that our lower bound
is tighter than the previous \cite{JJR}.
This supports  that the communicators in quantum key distribution can improve the security bounds by employing our derived result. {{Additionally,
our results}} will bring on  more potential applications in the further quantum communication. For examples, in the monogamy game, our new lower bound
can more precisely capture the tradeoff of entanglement monogamy during quantum information processing,
which {{enhances}} the precision of prediction of measurement outcome \cite{Pjc}. It means that the new tripartite uncertainty
relation should have an important application in precision measurements.
For another application, the new tripartite uncertainty relation implies the tighter security bounds on the ability that
the eavesdropper Eve predicts the measurement outcome of $Z$ measurement on  subsystem $A$ in the case where the additional term $\Delta>0$.
Furthermore, the bound of quantum secret key rate can be strengthened by utilizing the new lower bound of tripartite uncertainty relation.
In this sense, even though in the case that the eavesdropper Eve overhears the measurement outcomes in the communication of
Alice and Bob, a secure key may still be generated between them. Thereby, our security argument can effectively improve the security of quantum key distribution schemes.

\begin{acknowledgements}
This work was supported by the National Natural Science Foundation of China (Grant Nos. 61601002, 11875167 and 11575001),
Anhui Provincial Natural Science Foundation (Grant No. 1508085QF139), and the fund from CAS Key
Laboratory of Quantum Information (Grant No. KQI201701).
\end{acknowledgements}

\end{document}